\documentstyle[twocolumn,aps,epsf]{revtex}
\begin{document}
\title{Ferromagnetic insulating state in manganites: $^{55}$Mn NMR study}
\author{M. M. Savosta, V. I. Kamenev, V. A. Borodin}
\address{Donetsk Institute of Physics
\& Technics, Academy of Sciences of Ukraine, Rozy Luxembourg 72,
83114 Donetsk, Ukraine}
\author{P. Nov\'ak, M. Mary\v{s}ko, J. Hejtm\'anek}
\address{Institute of Physics,
Academy of Sciences of the Czech Republic, Na Slovance 2,
182 21 Praha 8, Czech Republic}
\author{K. D\"orr, M. Sahana}
\address{Institute of Solid State and Materials Research, IFW Dresden, P.O.B.
270016, D-01171 Dresden, Germany}
\author{A. Shames}
\address{Department of Physics, Ben-Gurion University of the Negev, P.O. Box
653, 84105, Be'er-Sheva, Israel}
\date{\today}
\maketitle
\begin{abstract}
$^{55}$Mn NMR was used to study the ferromagnetic insulating state in four different
manganites. In all cases the coexistence of two types of regions possessing
different NMR spectra and nuclear spin dynamics was found. In the ferromagnetic
insulating clusters the hopping frequency $f_{hop}$ of the electron holes is slower
comparing to the NMR frequency $f_{res}$ and two lines ascribed to 3+ and 4+ valence state
of Mn are observed. The relaxation rate increases rapidly with the increasing
temperature, so that these spectra can no longer be detected for temperatures
higher than $\approx$60 K. In ferromagnetic metallic clusters $f_{res} < f_{hop}$
and the motionally narrowed spectrum is observed in
a broad temperature interval, as the relaxation remains only moderately fast.
\end{abstract}

\section{Introduction}
Hole-doped manganites La$_{1-x}$A$_x$MnO$_3$ (A=Ca, Sr, Ba, Pb) have attracted
much attention in recent years. The reason is not only the 'colossal'
magnetoresistance that occurs in these systems, but also a rich variety of the
physical phenomena including intrinsically inhomogeneous ground states, phase
separation, charge/orbital ordering etc. \cite{dagotto}. The basic mechanism
that couples the spin and charge in these materials and produces the
ferromagnetic metallic-like (FMM) state is generally assumed to be described
by the double
exchange (DE) model. In this model, an electron hole tends to hop more easily
between the two successive Mn sites if both Mn site moments are aligned. Rather
intriguing is the existence of ferromagnetic but insulating (FMI) state, which
is beyond the DE concept. FMI state may be realized in several ways, in
particular it occurs when the doping is low, in the systems with strong
disorder or distortions in the lattice or for certain substitutions on the Mn
sites. The microscopic nature of these FMI states is far from being
clear - several models, including charge and/or orbital ordering, superexchange
ferromagnetic interactions, cluster-glass state, phase
segregation between ferromagnetic metallic and insulating
charge-ordered clusters, have been examined in this context
\cite{yamada,endoh,martinez,ghivelder,zhou}.

NMR is a suitable tool to provide further insight in the nature of the FMI state
as it probes locally the magnetic states and their dynamics. Ultraslow diffusion
of the small Jahn-Teller (JT) polarons in several insulating LaMnO$_{3+\delta}$
and La$_{1-y}$Ca$_{y}$MnO$_{3+\delta}$ manganites, gives rise to an
inhomogeneous loss and final disappearance
of the $^{139}$La NMR signal when the temperature
increases, as reported recently by Allodi {\it et al.} \cite{allodi}. This
'wipeout' of the $^{139}$La NMR signal intensity in the FMI
La$_{1-x}$Ca$_{x}$MnO$_{3}$ was observed also by Papavassiliou {\it et al.}
 \cite{papa} and it was associated with
a low frequency dynamics of the Mn octant cells. The $^{55}$Mn NMR in the
corresponding systems was studied in several papers
\cite{kapusta,kapusta1,belesi,allodi1}, the dynamics of the $^{55}$Mn
nuclear spins,
which is definitely an important item, was not addressed until now, however.

In this paper we present the results of detailed studies of the NMR spectra and
relaxation on the $^{55}$Mn nuclei, completed in some cases by $^{139}$La NMR,
for several FMI manganites, including two different self-doped
La$_{1-x}$Mn$_{1-y}$O$_3$,
low-doped La$_{0.84}$Sr$_{0.16}$MnO$_3$, and Ti-substituted
La$_{0.6}$Pb$_{0.4}$Mn$_{0.9}$Ti$_{0.1}$O$_3$ compounds. Two types of regions
associated with the FMM and FMI clusters, possessing quite different nuclear spin
dynamics and $^{55}$Mn NMR spectra, are detected for all compounds studied. The
NMR spectra arising from the FMM clusters consist of a single motionally
narrowed line, which is detected in a relatively broad temperature
interval. The exception is La$_{0.84}$Sr$_{0.16}$MnO$_3$ where more motionally
narrowed lines are observed below the metal-insulator transition.
The spin fluctuations on $^{55}$Mn
nuclei in FMI clusters are in a slow fluctuation limit, i.e. $f_{hop}<f_{res}$.
Corresponding spectra consist of two lines ascribed to Mn$^{4+}$ and Mn$^{3+}$
ions, and they can be detected for all compounds studied only for $T<$ 60 K.
This behavior is in line with the recovery of the signal amplitude on the
$^{139}$La nuclei in the same temperature interval. Two scenarios of the FMM
to FMI transition depending on the different ratio of FMM and FMI clusters
as well as possible nature of these clusters are discussed.

\section{Experimental}
\begin{figure}
\epsfxsize=6cm
\centerline{\epsfbox{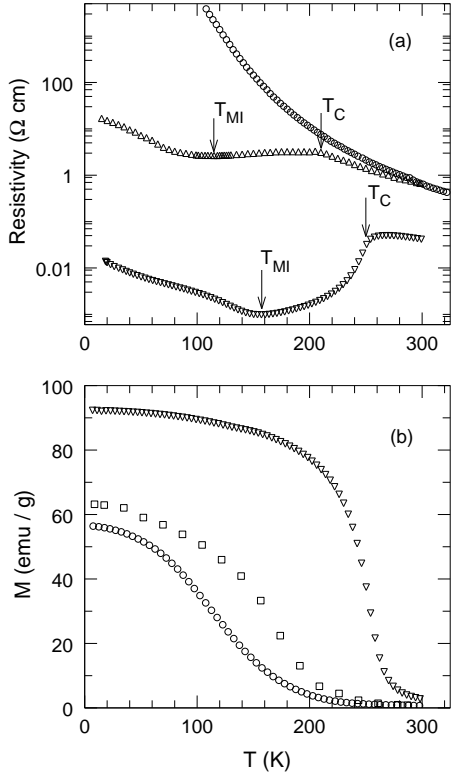}}
\caption{Temperature dependence of the resistivity (a) and magnetization in an
applied field of 0.5 T (b) for La$_{1-\delta}$MnO$_3$ ($\circ$),
La$_{0.94}$Mn$_{0.98}$O$_3$ ($\triangle$), and La$_{0.84}$Sr$_{0.16}$MnO$_3$
($\nabla$). Magnetization as a function of temperature for
La$_{0.6}$Pb$_{0.4}$Mn$_{0.9}$Ti$_{0.1}$O$_3$ measured in
0.3 T (lower panel, $\Box$).}
\end{figure}
The single crystal of La$_{0.94}$Mn$_{0.98}$O$_3$ and polycrystalline compound
La$_{0.6}$Pb$_{0.4}$Mn$_{0.9}$Ti$_{0.1}$O$_3$ used in the present study were
described in Refs. \cite{markovich,sahana}. The single crystal
La$_{0.84}$Sr$_{0.16}$MnO$_3$ was grown by the floating zone method. The self-doped
polycrystalline La$_{1-\delta}$MnO$_3$ ($\delta \approx$0.05) was prepared by standard solid
state reaction at 1000-1100$^\circ$C for 50 hours with intermediate grinding and
pressing. The temperature dependence of the electrical resistivity, measured
with the standard four-probe method, and magnetization curves are shown in Figs.
1a,b. The resistivity of La$_{0.84}$Sr$_{0.16}$MnO$_3$ exhibits a metallic-like
behavior below $T_C$= 250 K. As temperature decreases further, the curve shows an
upturn in the ferromagnetic phase below the metal-insulator transition
temperature $T_{MI}=$157 K. The FMI state ($d\rho/dT<0$) is realized below
$T_{MI}$. A similar behavior is observed for La$_{0.94}$Mn$_{0.98}$O$_3$ with
$T_C$= 210 K \cite{markovich} and $T_{MI}=$115 K,
though the value of the resistivity is by one to
three orders of magnitude higher then that of La$_{0.84}$Sr$_{0.16}$MnO$_3$. The
self-doped La$_{1-\delta}$MnO$_3$ is an insulator, the onset of the magnetic
ordering is at about 190 K. The resistivity of
La$_{0.6}$Pb$_{0.4}$Mn$_{0.9}$Ti$_{0.1}$O$_3$ ($T_C$=180 K) was not measured, but
it is known that such level of Ti doping on the Mn site leads to an insulating
behavior for La$_{0.7}$Ca$_{0.3}$Mn$_{1-x}$Ti$_{x}$O$_3$ \cite{liu,cao}.

The NMR spectra were recorded by a two-pulse spin-echo method at temperatures
between 22 and 250 K using a noncoherent home-build spectrometer with frequency
sweep and boxcar detector signal averaging. The NMR spectra were obtained by
measuring the integrated intensity of the spin-echo versus frequency using
$\tau$=3-4 $\mu$s, where $\tau$ is the time separation between the two pulses. The
spin-spin relaxation was studied by measuring the form of the spin-echo decay as a
function of $\tau$. All the signals detected come from the regions with strong
ferromagnetic correlations as follows from the characteristic high values of the
enhancement factor $\eta \approx$500-2500. The spin-echo amplitudes were
corrected for the $f^2$-type frequency response and the $1/T$ reduction, that
results from the Curie law for the nuclear magnetic moment.

\begin{figure}
\epsfxsize=5.8cm
\centerline{\epsfbox{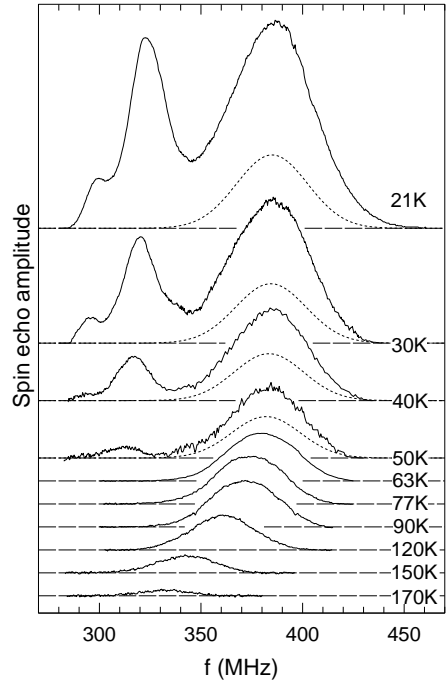}}
\caption{La$_{1-\delta}$MnO$_3$. $^{55}$Mn NMR spectra taken at ten different
temperatures. The dotted curves correspond to the contribution of the
FMM clusters to the spectra.}
\end{figure}

\section{Results and discussion}
The evolution of the $^{55}$Mn NMR spectra in La$_{1-\delta}$MnO$_3$ with the
temperature is displayed in Fig. 2. For $T\geq$ 63 K the NMR spectra consist of
a single line similarly as found in conventional FMM manganites
\cite{matsumoto}. This line is attributed to a motional narrowing - the electron
holes hop over the manganese sites with a rate faster then the frequency
splitting $\Delta f$ of the Mn$^{3+}$ and Mn$^{4+}$ resonances, and as a result all
$^{55}$Mn nuclei feel the same averaged hyperfine field. The reduction of the
spin-echo amplitude of this line with increasing $T$ is mainly connected with
the shortening of the spin-spin relaxation time $T_2$, which becomes comparable
to the time separation $\tau$. Below 63 K the NMR spectrum qualitatively changes,
as new contributions gradually develop. The line at around 320 MHz corresponds to
the Mn$^{4+}$ resonance \cite{matsumoto}. The satellite line at around 300 MHz,
which may be attributed to a Mn$^{4+}$ resonance for specific Mn sites, is
observed only in this sample and it will be discussed later. More important, the
amplitude of the signal at about 390 MHz increases substantially below 63 K,
indicating additional contribution in this frequency region too.

\begin{figure}
\epsfxsize=7cm
\centerline{\epsfbox{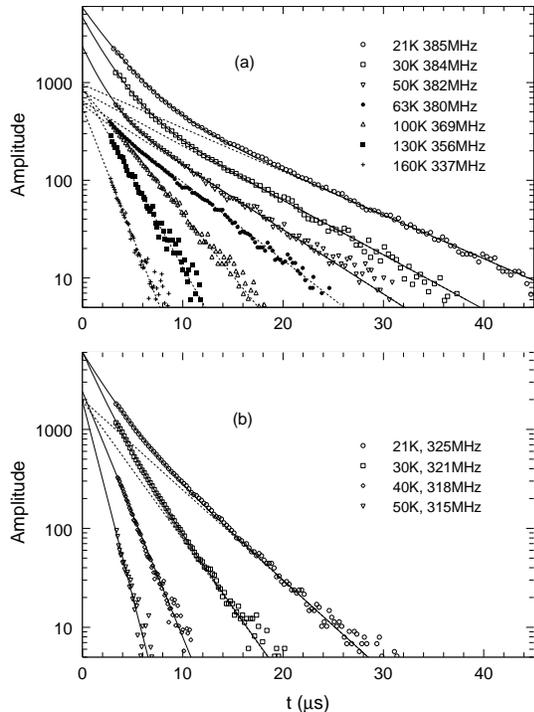}}
\caption{La$_{1-\delta}$MnO$_3$. Spin-echo decay at the maximum of the
FMM line (a) and Mn$^{4+}$ line (b). The solid curves correspond to the
two-exponential fit, the dotted curves correspond to the slower relaxing
component.}
\end{figure}

The spin-echo decays taken at the maximum of the motionally narrowed
FMM line
for different temperatures reveal this contribution (Fig. 3a). For $T\geq$ 63 K
the decays are exponential, $A=A_0\exp(-2\tau/T_2)$. The slope of the decays
changes monotonically in the whole temperature range (taking the tail ends of
the spin-echo decay curves for low $T$), reflecting the temperature dependence
of $T_2$ for the FMM line. Moreover, the amplitude $A_0$ of the
FMM line,
extrapolated to $\tau$=0 (dashed lines in Fig. 3) is nearly constant in the limits of
the experimental error $\pm$20\%. The contribution of FMM line to the spectra at
low temperatures is schematically shown in Fig. 2 by dotted curves. From the
data in Fig. 3a it is clearly seen that the increase of the signal intensity
below 63 K is due to an additional, fast relaxing contribution to the spin-echo
decays. The possibility that this contribution originates from the domain walls
can be safely ruled out as it follows from the form of the spin-echo decays taken
at different values of radiofrequency (rf) power (Fig. 4a). It is seen that the
optimal value of the rf field for the fast relaxing component is four times
larger compared to the one for FMM line. Thus, faster relaxing regions possess
smaller NMR enhancement contrary to expected behavior of the NMR signal from
the domain walls. The spin-spin relaxation of Mn$^{4+}$ line is strongly
temperature dependent, corresponding spin-echo decay curves are not of a
single-exponential type (Fig. 3b). The form of the spin-echo decay is rather
insensitive to the change of the rf field around its optimal value in this case
(Fig. 4b), indicating certain distribution of relaxation times in the regions with
the similar magnetocrystalline anisotropy.

\begin{figure}
\epsfxsize=6.7cm
\centerline{\epsfbox{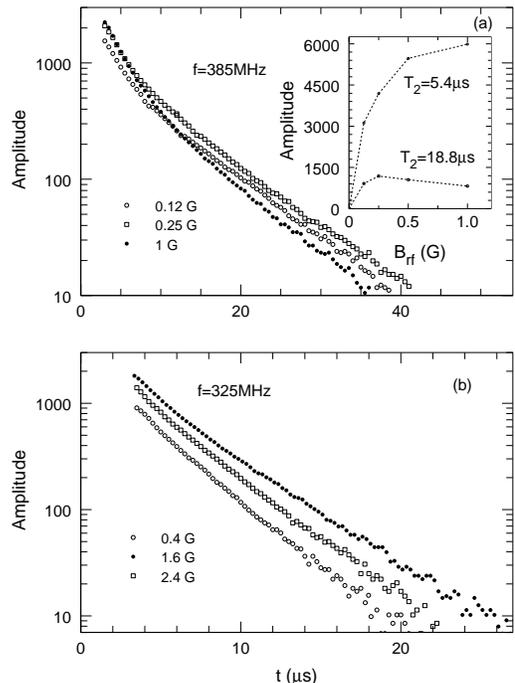}}
\caption{La$_{1-\delta}$MnO$_3$. (a) Spin-echo decay
at the maximum of the FMM line
for three values of the rf power. In the inset the dependence of the
spin-echo amplitude on the rf power is shown for slower relaxing component
($T_2$=18.8 $\mu$s)
and faster relaxing component
($T_2$=5.4 $\mu$s) of the line. (b) Spin-echo decay at the maximum of the
Mn$^{4+}$ line for three values of the
rf field around its optimal value.}
\end{figure}

The $^{55}$Mn NMR suggests  that in
La$_{1-\delta}$MnO$_3$
two different types of ferromagnetic regions exist.
In what follows these regions are called FMM and FMI clusters. The signal
arising from the FMM clusters consists of a single motionally narrowed line,
which is detected in relatively broad temperature interval thanks to the
fact that corresponding spin-spin relaxation is only moderately
fast. The FMI clusters give rise to two lines at about 320
and 390 MHz which can be detected only for $T\leq$ 60 K because of the
inhomogeneous and very strongly temperature dependent spin-spin relaxation. As
we show below this picture is typical for all compounds studied. It is clear
from Fig. 3 that the spin-spin relaxation strongly influences the amplitude
of the NMR spectra
even if the shortest $\tau$ is used. To make the NMR spectra of remaining three
compounds more representative, we tried to correct the line's
amplitude for the measured spin-spin relaxation. The fits of corresponding spin-echo
decays by two or single exponential were employed. We checked that a
stretched exponential form of the fitting function,
often used in the literature,
would not affect our estimate of the temperature dependence of $A_0$.

\begin{figure}
\epsfxsize=6.5cm
\centerline{\epsfbox{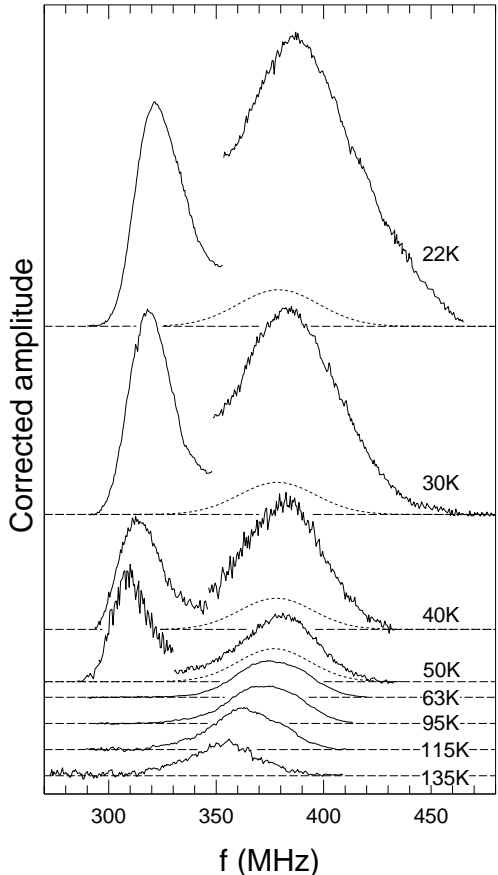}}
\caption{La$_{0.6}$Pb$_{0.4}$Mn$_{0.9}$Ti$_{0.1}$O$_3$. Corrected $^{55}$Mn NMR spectra
at several temperatures. The dotted curves correspond to the contribution of the
FMM clusters to the spectra.}
\end{figure}

Corrected NMR spectra of La$_{0.6}$Pb$_{0.4}$Mn$_{0.9}$Ti$_{0.1}$O$_3$,
La$_{0.94}$Mn$_{0.98}$O$_3$, and La$_{0.84}$Sr$_{0.16}$MnO$_3$ are presented
in Figs. 5, 7, 8, respectively. The correction is qualitative only as for
low and
high frequency peaks we used the decay of the spin-echo signal at the
corresponding amplitude maxima and neglected the fact that it changes
continuously with the frequency. As a result a discontinuity appears in the
corrected spectra. For La$_{0.6}$Pb$_{0.4}$Mn$_{0.9}$Ti$_{0.1}$O$_3$, the
situation is analogous to that discussed above for La$_{1-\delta}$MnO$_3$ except
that the NMR enhancement was found to be only slightly larger for FMM compared
to FMI clusters. This fact leads to a more pronounced contribution of the fast
relaxing component to the spin-echo decay curves (Fig. 6), though the relative
volume of the FMM clusters in the Ti-substituted sample, estimated from the
$^{55}$Mn NMR, is about 10\% of the total volume, while for
La$_{1-\delta}$MnO$_3$ it is only 1.5-3\%. For both compounds the FMM volume
practically do not change with the temperature. The relaxation for the FMI signal is
very strongly temperature dependent and evidently there exists a distribution of
the relaxation rates. As a consequence when relaxation rate increases,
increasing part of the nuclear spins relaxes before it could be detected. This
is the reason for gradual decrease and final disappearance of the FMI NMR
signal, though they were corrected (Fig. 5). The number of Mn spins in the FMI
clusters is in fact preserved as the magnetic moment in the corresponding
temperature interval changes only slightly (Fig. 1b).

\begin{figure}
\epsfxsize=8cm
\centerline{\epsfbox{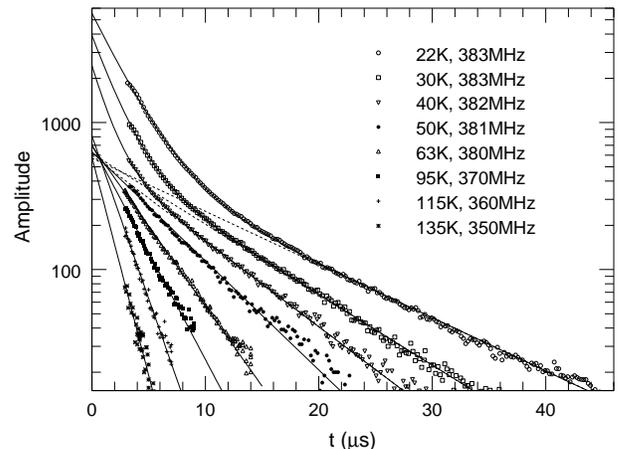}}
\caption{La$_{0.6}$Pb$_{0.4}$Mn$_{0.9}$Ti$_{0.1}$O$_3$. Spin-echo decay taken at the maximum of the FMM line.}
\end{figure}

A metal-insulator transition occurs in La$_{0.94}$Mn$_{0.98}$O$_3$ compound ($T_{MI}$=115 K) and in
La$_{0.84}$Sr$_{0.16}$MnO$_3$ ($T_{MI}$=157 K) and this is reflected in the corresponding NMR spectra. For
La$_{0.94}$Mn$_{0.98}$O$_3$ (Fig. 7) the amplitude of the FMM signal decreases considerably below
$\sim$150 K pointing to the reduction of the summary volume of the FMM clusters
around the MI transition (see also Fig. 10b). Below $\sim$60 K
the FMI
clusters are detected by NMR,
their behavior being similar to the one described above. A quite different
situation is found in the La$_{0.84}$Sr$_{0.16}$MnO$_3$ compound. Here the FMM regions give the
dominating contribution to the NMR spectrum in the whole temperature interval
studied. However, below the MI transition the FMM NMR spectrum no longer
consists of a single gaussian line as an obvious tail starts to develop on the low
frequency side of the line. As discussed in detail elsewhere \cite{savosta0}
this corresponds to a charge and/or orbital ordering in the FMM regions.
Analogously to other three compounds a FMI signal can only  be detected
below $\sim$60 K, but because of the dominance of FMM signal, only the
Mn$^{4+}$ resonance can be distinguished unambiguously.

\begin{figure}
\epsfxsize=7.5cm
\centerline{\epsfbox{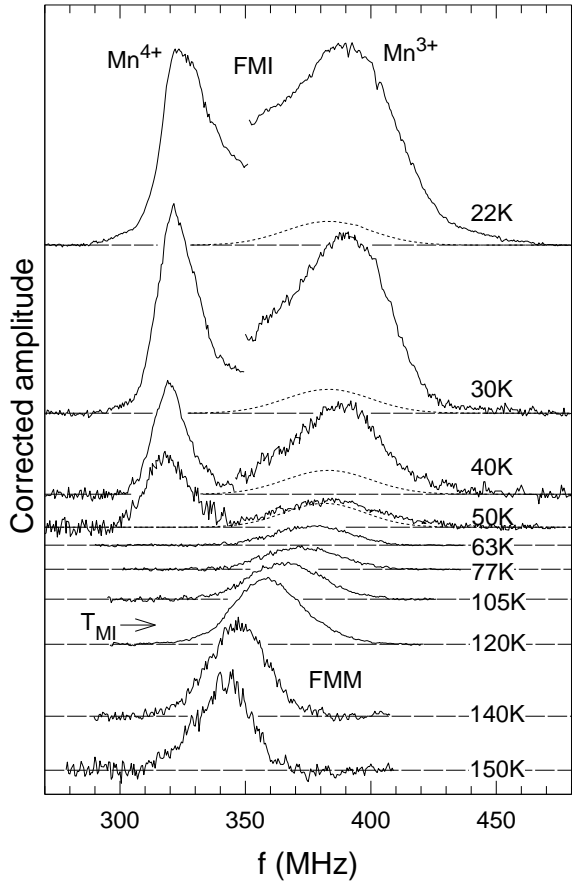}}
\caption{La$_{0.94}$Mn$_{0.98}$O$_3$. Corrected $^{55}$Mn NMR spectra at several temperatures.
The dotted curves correspond to the contribution of the FMM clusters to the
spectra.}
\end{figure}

\begin{figure}
\epsfxsize=7.5cm
\centerline{\epsfbox{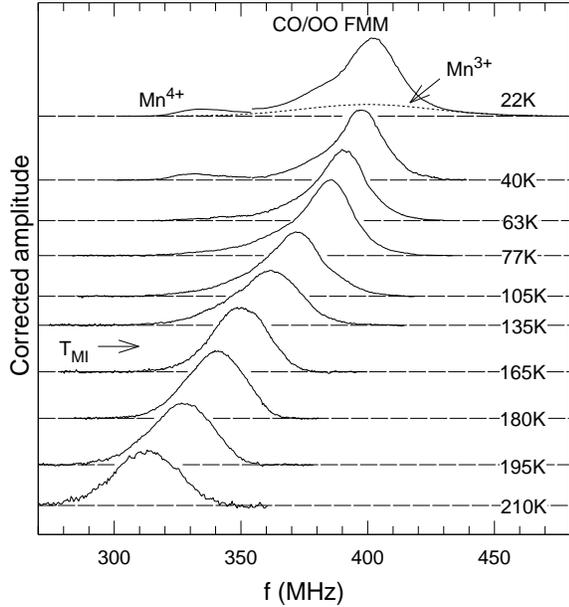}}
\caption{La$_{0.84}$Sr$_{0.16}$MnO$_3$. Corrected $^{55}$Mn NMR spectra taken
at several temperatures. The dotted line at the 22 K shows schematically the
contribution of the Mn$^{3+}$ ions.}
\end{figure}

\begin{figure}
\epsfxsize=6.1cm
\centerline{\epsfbox{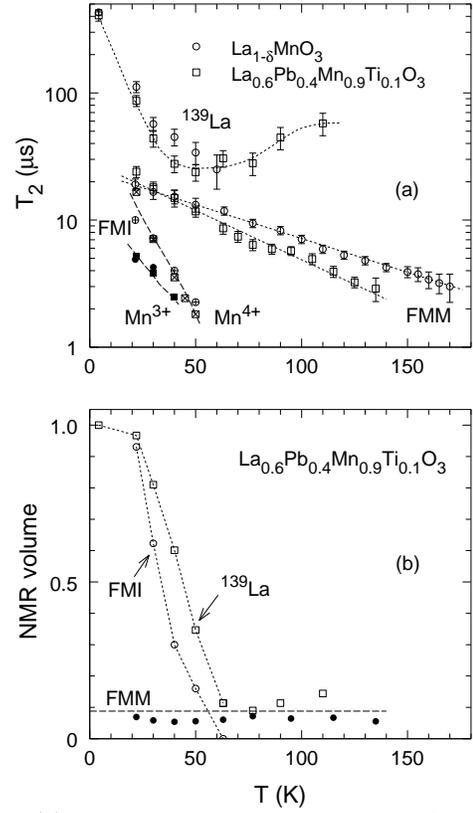}}
\caption{(a) Temperature dependence of the spin-spin relaxation time $T_2$
for $^{55}$Mn nuclei in the FMM and FMI clusters and for $^{139}$La nuclei.
(b) Temperature dependence of the relative reduced volume of the FMM ($\bullet$)
and FMI ($\circ$) clusters determined from the areas under corresponding
$^{55}$Mn NMR signals. $\Box$ denote the reduced volume determined from
the area under the $^{139}$La signal.}
\end{figure}

The results of the analysis of the La$_{1-\delta}$MnO$_3$ and La$_{0.6}$Pb$_{0.4}$Mn$_{0.9}$Ti$_{0.1}$O$_3$ NMR spectra are displayed in
Fig. 9. To obtain more complete information the $^{139}$La NMR was also studied
for these compounds. The spin-spin relaxation on the La nuclei is strongly
temperature dependent and it also becomes strongly inhomogeneous when the
temperature increases, similarly as reported in \cite{allodi,papa}. Shown in
Fig. 9a are the values of $T_2(T)$ for $^{55}$Mn in FMI and FMM clusters as well
as for the $^{139}$La where, as shown below, the contribution from the
FMI clusters dominates at low temperatures, while at higher temperatures only
the contribution from the FMM clusters remains. For Mn$^{4+}$ and La, where the
relaxation is nonexponential, the values of $T_2$ determined from the tail ends
of the spin-echo decay are presented. As seen from Fig. 3b this provides an
appropriate characteristic of the evolution of the spin-echo decay with the
temperature.

\begin{figure}
\epsfxsize=7.5cm
\centerline{\epsfbox{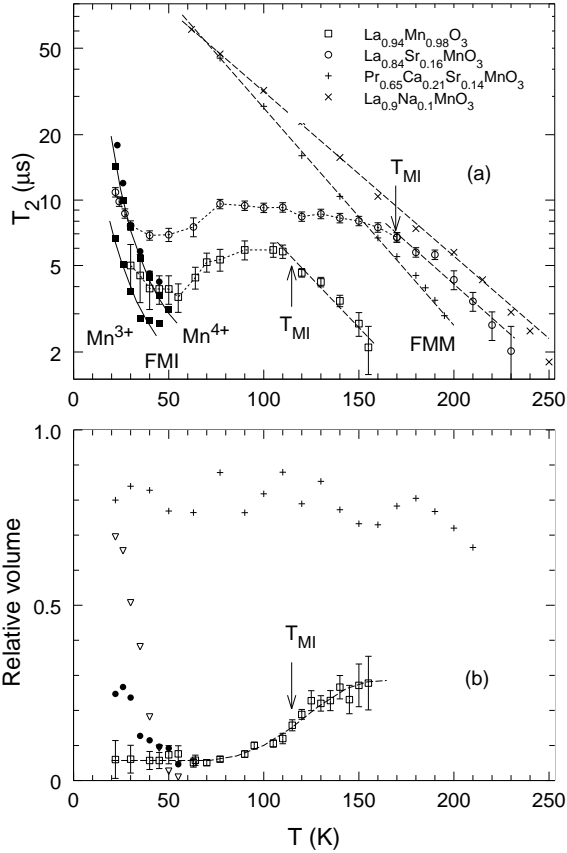}}
\caption{(a) Temperature dependence of the spin-spin relaxation time $T_2$
for $^{55}$Mn nuclei in the FMM and FMI clusters. For comparison the
data for two metallic-like manganites La$_{0.9}$Mn$_{0.1}$MnO$_3$
($T_C$=266 K) and Pr$_{0.65}$Ca$_{0.21}$Sr$_{0.14}$MnO$_3$ ($T_C$= 205 K)
are also displayed.
(b) Temperature dependence of the relative volumes of the FMM and FMI
clusters.  For La$_{0.94}$Mn$_{0.98}$O$_3$ the volumes of the FMM clusters
are denoted by $\Box$, the volume of the FMI clusters determined
from the Mn$^{4+}$ and Mn$^{3+}$ line are denoted by $\bullet$ and $\nabla$,
respectively. For La$_{0.84}$Sr$_{0.16}$MnO$_3$ only the FMM volume
is displayed (+).}
\end{figure}

The form of temperature dependence of $T_2$ in the FMM clusters is similar to
the one found in the metallic-like manganites \cite{savosta99}, though the
the relaxation is considerably faster. The $^{55}$Mn
relaxation in the FMI clusters increases rapidly with the temperature, its
behavior below 60 K being remarkably similar to that of $^{139}$La. This fact
strongly suggests that in the FMI clusters
the origin of the relaxation is the same for both nuclei.
In the metallic-like manganites the $^{55}$Mn relaxation is
due to the fluctation of the hyperfine field induced by the hopping of the
electron holes. This mechanism is, however, ineffective for the La nuclei and,
as a consequence, the temperature dependence of $T_2$ for La is flat and quite
different from that of the Mn nuclei in this case \cite{savosta99}. The interconnection of
the $^{55}$Mn and $^{139}$La relaxation in the FMI regions may be then
understood as follows - the hopping of the holes is much slower comparing to the
metallic manganites and the charge carriers may be represented as the JT small
polarons, their movement being accompanied by a lattice excitation leading to a
fluctuation of the electric field gradient (EFG) on the La sites.
Thus both the fluctuation of the EFG
on the La sites and the
fluctuation of the EFG and the hyperfine field on the Mn sites have the
same characteristic
time $\tau_{JT}$ connected with the movement of the JT polaron which leads to similar
temperature dependence of corresponding relaxations. Note that for the
$^{139}$La relaxation this mechanism was recently proposed by Allodi {\it et al.}
\cite{allodi}. These authors showed that at temperatures below 60 K the
$^{139}$La NMR in FMI manganites corresponds to the 'slow fluctuation' limit
i.e. $\tau_{JT}\;\gamma_n B_{hf} > 1$. Therefore, the two lines we observe in the
$^{55}$Mn NMR spectra of the FMI clusters correspond to 3+ and 4+ valence states of
the Mn ion, which is the same conclusion as reached in \cite{allodi1}.  Note
that this situation corresponds to a slow hopping, rather then to a
localization of the charge
carriers as for the localized Mn$^{3+}$ ions the NMR spectrum is
strongly anisotropic and the nuclear spin-lattice relaxation time $T_1$ is
very short. In this situation the spin-spin relaxation
is governed by the spin-lattice channel
with a limiting relation $T_2$=2$T_1$. In contrast, down to 22 K the spin-lattice
relaxation for both Mn$^{4+}$ and Mn$^{3+}$ lines was checked to be relatively slow
with a ratio $T_1$/$T_2$ $\approx$30-60 in all compounds in question.
Rather surprising proximity of $f_{res}$
Mn$^{3+}$ in the FMI clusters and $f_{res}$ of the motionally narrowed line in
the FMM clusters could be connected with different local distortions in these two
types of clusters.

\begin{figure}
\epsfxsize=8cm
\centerline{\epsfbox{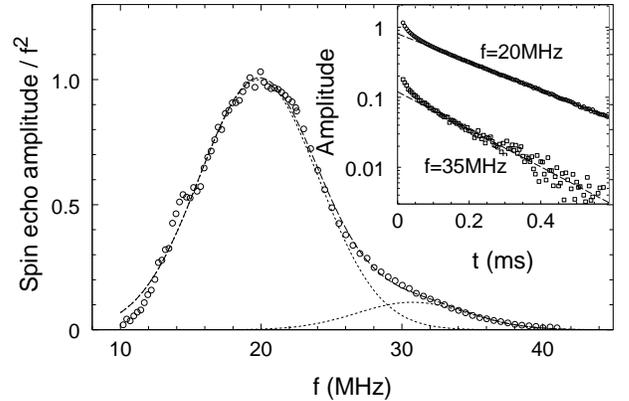}}
\caption{La$_{1-\delta}$MnO$_3$. $^{139}$La NMR spectrum at 4.2 K and its decomposition. In
the inset the spin-echo decay at two frequencies around the maxima of
the two components of the spectrum are shown. }
\end{figure}

The temperature dependence of the normalized amplitudes of the $^{55}$Mn signal
from FMI (summing the Mn$^{3+}$ and Mn$^{4+}$ contributions) and FMM clusters
are compared with the normalized amplitude of the $^{139}$La
signal which was corrected for observed $T_2$ in Fig. 9b.
It is clearly seen that above 60 K the La signal comes from the FMM clusters,
while at lower temperatures its increase is connected with the contribution from
the FMI regions. Note that the recovery of the signal from the FMI regions has
a similar character for both Mn and La nuclei.

The results of the analysis of $^{55}$Mn NMR spectra in La$_{0.94}$Mn$_{0.98}$O$_3$ and
 La$_{0.84}$Sr$_{0.16}$MnO$_3$ are
displayed in Fig. 10. The NMR relaxation in the FMI clusters is analogous to
that described above, while its behavior in the FMM clusters is more complex and
intriguing. Above the metal-insulator transition the temperature dependence
of the relaxation is
exponential-like as in metallic manganites though for La$_{0.94}$Mn$_{0.98}$O$_3$ the
magnitude of $T_2$ is considerably smaller, below $T_{MI}$ it becomes more
flat for both compounds exhibiting a minimum at $\sim$50 K (Fig. 10a). This can
be naturally connected to the slow fluctuations which in the same temperature
region lead to the disappearance of the NMR signal in the FMI clusters.
Such attribution of the minima in $T_2$ implies that there is a hyperfine coupling
between the JT small polarons existing in the FMI clusters
 and $^{55}$Mn nuclear spins in the FMM clusters, i.e.
the FMI and FMM systems are closely intertwined. It is interesting to note in
this context that for La$_{0.94}$Mn$_{0.98}$O$_3$ the decrease of the volume of FMM clusters (Fig.
10b) is accompanied by a reduction of the NMR enhancement factor approx. 1.7 times
between 120 and 77 K, and at lower temperatures it becomes close to
the NMR enhancement for FMI clusters.
On the other hand, for La$_{1-\delta}$MnO$_3$ and La$_{0.6}$Pb$_{0.4}$Mn$_{0.9}$Ti$_{0.1}$O$_3$ samples no visible minima in
$T_2$ for FMM signal is observed (Fig. 9a). This may imply that for these
systems the minority FMM regions and the FMI matrix are less intimately
intertwinned.

As mentioned above, a satellite of the Mn$^{4+}$ line in the $^{55}$Mn spectrum
at $\sim$300 MHz is observed in La$_{1-\delta}$MnO$_3$ (Fig. 2). The splitting $\sim$25 MHz is
too large to be explained by the change of the dipolar or/and supertransferred
fields it could be, however, caused by a redistribution of the spin density
around specific Mn sites. Such redistribution, decreasing local spin density on
the Mn site would lead to an increase of the transferred hyperfine interaction
on the nearest La sites due to the enhanced Mn-O covalency. Indeed an unresolved
satellite line on the high frequency side of the $^{139}$La spectrum is detected
in this compound as seen in Fig. 11. Interestingly a similar high frequency tail
of the La NMR was observed in the (LaCa) manganites
\cite{papa,yoshinari}, while a satellite in the $^{55}$Mn spectra could be found
in the results reported for these compounds and for a self-doped LaMnO$_3$
\cite{kapusta1}.

Let us present now a general picture of FMI state in manganites which can be
inferred from the above described
results. Perturbations in these systems hinder the motion of the charge
carriers, reducing thus the
strength of the DE interaction. Slower hopping of the electron holes
increases the role of their interaction with the lattice and, as a result, the
clusters where charge carriers can be viewed as JT small polarons grow at the
expense of the metallic-like host. The resulting state may be ferromagnetic
insulating if the polaronic regions dominate (La$_{1-\delta}$MnO$_3$,
La$_{0.6}$Pb$_{0.4}$Mn$_{0.9}$Ti$_{0.1}$O$_3$), or the
transition to metallic like conductivity with relatively high resistivity may
occur through the percolation of the FMM clusters as their
total volume increases with increasing
temperature (La$_{0.94}$Mn$_{0.98}$O$_3$). A specific case represents the system
La$_{0.84}$Sr$_{0.16}$MnO$_3$, where the insulator to metal transition in the
FM state is due to the
breakdown of charge and/or orbital ordering in the FMM regions. Accordingly, the
resitivity in ferromagnetic metallic state is closer to the one in conventional
manganites in this case. The minority polaronic regions are also detected for
La$_{0.84}$Sr$_{0.16}$MnO$_3$ and we expect that their role will increase for
the lower doping. This
may explain the different behavior of FMI state in La$_{1-x}$Sr$_{x}$MnO$_3$ for
slightly different $x$, in particular
for 0.1$\leq$x$\leq$0.14 the FMI state is stabilized under
pressure and field \cite{martinez,zhou}, while for $x$=0.15, 0.16 the FMI state
is suppressed by pressure \cite{zhou,moritomo}. Further NMR experiments are
required to clarify the problem.

The effect of the inhomogeneity on the motion of the charge carriers,
that is the distribution of the activation
energies of JT polaron hopping seems to be similar for all cases, leading to
disappearance of the signals on Mn nuclei at $T\sim$60 K. An analogous behavior was
found in La$_{0.825}$Ca$_{0.175}$MnO$_3$ \cite{belesi}. Also Mn$^{4+}$
resonance in La$_{0.9}$Ca$_{0.1}$MnO$_3$ was detected up to 77 K only
\cite{kapusta1}. Further, the disappearance or reduction of the amplitude of
signal on the La nuclei was reported for a number of self-doped and
(LaCa) manganites at
the temperature range 50-100 K in Refs. \cite{allodi,papa} and it is the same
behavior as detected for La$_{1-\delta}$MnO$_3$ and
La$_{0.6}$Pb$_{0.4}$Mn$_{0.9}$Ti$_{0.1}$O$_3$ samples in the present study.
All these
facts suggest that the polaronic state is an intrinsic response of the DE system
to the perturbation rather then being specific for a given type of disorder.
Thus at least for the FMI
system with $T_C\sim$150-200 K the ferromagnetic state is still controlled
by DE through thermally activated hopping of small JT polarons. On the other
hand even for strongly disordered
La$_{1-\delta}$MnO$_3$ a small fraction of the fast hopping
holes survives in the polaronic host. This rises the question whether the JT
polarons can alone produce the long range ferromagnetic ordering, or a small
fraction of the fast hopping holes is an essential component of the FMI state.

The NMR signals arising from the FMI clusters correspond to the slow fluctuation
limit which means the freezing of the polarons on a time scale
comparable to the duration of a NMR spin-echo experiment. Different mechanisms
including quasistatical orbital \cite{papa} and charge ordering \cite{allodi}
have been suggested in the literature in order to explain the corresponding
ground magnetic state. The quasistatical orbital hypothesis is based on observation
of a high frequency tail in $^{139}$La NMR spectra of (LaCa) manganites at low
temperatures. We also found such a feature in the La$_{1-\delta}$MnO$_3$
but it is absent in the La$_{0.6}$Pb$_{0.4}$Mn$_{0.9}$Ti$_{0.1}$O$_3$ spectra.
Also the corresponding satellite line in $^{55}$Mn spectra is detected only for
La$_{1-\delta}$MnO$_3$.
We conclude therefore that this can hardly serve as a characteristic feature of
the FMI state. Moreover even a short range orbital ordering seems to be
unlikely in the case of the strongly disordered
La$_{1-\delta}$MnO$_3$.
An alternative possibility may be a cluster-glass state. Such a state has been
detected by specific heat and magnetic measurements in self-doped manganites
LaMnO$_{3+\delta}$ \cite{ghivelder}. In particular, authors of Ref.
\cite{ghivelder} observed low temperature peak of the imaginary component of
susceptibility, the position of which strongly depends on the measuring
frequency. Shown in Fig. 12 is the ac susceptibility of La$_{1-\delta}$MnO$_3$
sample. The frequency dependence of the real and
imaginary part of the susceptibility as well as their absolute values
are indicative of the cluster glass behaviour similarly as reported
 by Ghivelder {\it et al.}
The situation may be similar to that observed in the lanthanum cuprates, where
slowing down of the spin and/or charge fluctuations results in a strong peak in
the spin-lattice relaxation for the La nuclei and in a loss of signal on
the Cu nuclei and it
is ascribed to the glassy magnetic state \cite{julien,hunt,curro}. It is
interesting that for both Cu and La signals stretched exponential like spin-echo
decay is observed in the slow fluctuation limit similarly to the one for FMI
signals in manganites, while in the motionally
narrowed limit, below a transition temperature $T_{charge}$, corresponding to the
localization of charge-ordered stripes, a Lorentzian single-exponential
spin-echo decay is detected \cite{hunt,teitel'baum,hunt1}.

\begin{figure}
\epsfxsize=6cm
\centerline{\epsfbox{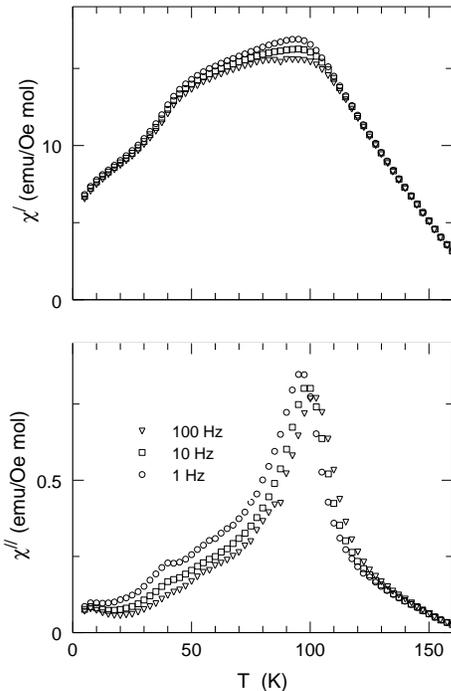}}
\caption{ Temperature dependence of the
 ac susceptibility for 1, 10 and 100 Hz for La$_{1-\delta}$MnO$_3$.}
\end{figure}

\section{Conclusions}
The study of $^{55}$Mn NMR in four manganites, in which the ferromagnetic
insulating state is caused by diverse reasons, showed that in all
cases the metallic-like (FMM, $f_{hop}>\Delta f_{res}$) and insulating
(FMI, $f_{hop}<\Delta f_{res}$) regions coexist, though the ratio of
their volumes differs considerably in different compounds. The dynamics of
the nuclear spins in the FMI clusters is almost independent of the
compound, pointing to similarity of the charge carrier dynamics, despite
the different origin of the FMI state in the manganites studied.
The temperature dependence of $^{139}$La and $^{55}$Mn relaxation in
the FMI clusters is analogous which can be understood as due to the
EFG fluctuation on the La nuclei and combined EFG and hyperfine field
fluctuation on Mn nuclei connected with the motion of the Jahn-Teller
polaron.
In two compounds with the metal-insulator transition the temperature
dependence of
the relaxation rate in FMM and FMI clusters indicates that these two
types of regions should be intertwinned on a microscopic scale.
In the self-doped La$_{0.94}$Mn$_{0.98}$O$_3$ the insulator to
metal transition occurs because of the percolation of the FMM clusters,
as their total volume increases with increasing temperature, while in
La$_{0.84}$Sr$_{0.16}$MnO$_3$ the reason is the breakdown of the
charge and/or orbital ordering.
Finally we suggest that the FMI state in manganites may be viewed as
a cluster-glass, which is supported by the character of the
temperature and frequency dependence of the ac susceptibility.

We are grateful to Y. Tomioka and to M. Greenblatt for providing
La$_{0.84}$Sr$_{0.16}$MnO$_3$ and La$_{0.94}$Mn$_{0.98}$O$_3$ single crystals,
respectively. We also thank to T. N. Tarasenko for sintering the
La$_{1-\delta}$MnO$_3$ polycrystal, to
V. D. Doroshev for measuring the resistivity of La$_{1-\delta}$MnO$_3$,
to J. Englich and
J. Kohout for providing the possibility of $^{139}$La NMR measurements
at liquid helium temperature at the coherent spectrometer at the
Charles University, Prague.
This work was supported by the grant
202/00/1601 of Grant Agency of the Czech Republic and
grant A1010202 of the Grant Agency of AS CR.

\end{document}